\documentclass[a4paper,11pt]{article}
\pdfoutput=1
\usepackage[utf8x]{inputenc}
\usepackage[margin=0.9in]{geometry}
\usepackage{amsmath}
\usepackage{latexsym}
\usepackage{amssymb}
\usepackage[english]{babel}
\usepackage{color}
\usepackage[multiple]{footmisc}
\usepackage{jheppub}
\pdfoutput=1
\usepackage{jheppub}
\usepackage{breakurl}
\usepackage{mathrsfs}
\usepackage{amsmath}
\usepackage{amsfonts}
\usepackage{mathrsfs}
\usepackage{slashed}
\usepackage{epigraph}
\usepackage{fancyhdr}
\usepackage{amssymb}
\usepackage{graphicx}
\usepackage{longtable}
\usepackage{makecell}
\usepackage{amscd}
\usepackage{tabu}
\usepackage{bm}
\usepackage{enumerate}
\usepackage{color}
\usepackage{hyperref} 
\hypersetup{linktocpage=true}
\usepackage[all]{hypcap}
\hypersetup{
   bookmarks=true,         
   unicode=false,          
   pdftoolbar=true,        
   pdfmenubar=true,        
   pdffitwindow=false,     
   pdfstartview={FitH},    
   pdftitle={My title},    
   pdfauthor={Author},     
   pdfsubject={Subject},   
   pdfcreator={Creator},   
   pdfproducer={Producer}, 
   pdfkeywords={keyword1} {key2} {key3}, 
   pdfnewwindow=true,      
   colorlinks=true,        
   linkcolor=blue,         
   citecolor=magenta,      
   filecolor=magenta,      
   urlcolor=cyan           
}
\usepackage{fancyhdr}
\usepackage{booktabs}

\preprint{Cavendish-HEP-14/14}
\title{Composite leptoquarks and anomalies in $B$-meson decays}
\author[a]{Ben Gripaios}
\author[a,b]{M. Nardecchia}
\author[b]{S. A. Renner}

\affiliation[a]{\normalfont{Cavendish Laboratory, University of Cambridge, 
J.J. Thomson Avenue, Cambridge, CB3 0HE, UK}}
\affiliation[b]{\normalfont{DAMTP, University of Cambridge, 
Wilberforce Road, Cambridge, CB3 0WA, UK}}

\emailAdd{gripaios@hep.phy.cam.ac.uk} 
\emailAdd{m.nardecchia@damtp.cam.ac.uk}
\emailAdd{sar67@cam.ac.uk} 

\abstract{
We attempt to explain recent anomalies in
semileptonic $B$ decays at LHCb via a
composite Higgs model, in which both the Higgs and an $SU(2)_L$-triplet
leptoquark arise as pseudo-Goldstone bosons of the strong
dynamics. Fermion masses are assumed to be generated via the mechanism
of partial compositeness, which largely determines the leptoquark
couplings and implies non-universal lepton interactions.
The latter are needed to accommodate tensions in the $b \to s \mu \mu$
dataset and to be consistent with a discrepancy measured at LHCb in
the ratio of $B^+ \to K^+ \mu^+ \mu^-$ to $B^+ \to K^+ e^+ e^-$
branching ratios. The data imply that the leptoquark should have a
mass of around a TeV. We find that the model is not in conflict with
current flavour or direct production bounds, but we identify a
few observables for which the new physics contributions are close to
current limits and where the leptoquark is likely to show up in future
measurements. The leptoquark will be pair-produced at the LHC and
decay predominantly to third-generation quarks and leptons, and LHC13
searches will provide further strong bounds.
}
\begin{document}
\maketitle

\section{Introduction}
The first run of the LHC brought us the long-awaited discovery of the
Higgs boson, but no firm evidence for the physics
beyond the Standard Model (SM) 
needed to avoid fine-tuning of the weak scale. This was
perhaps not unexpected, given the plethora of indirect constraints
on new physics coming from, {\em e.g.}, flavour
physics and electroweak precision tests. Typically these point at
scales of new physics way beyond a TeV; even when we invoke all
the dynamical dirty tricks that we know of, the best we can do is to
lower the possible scale of new physics to perhaps a few TeV.
Therefore, there seems to be, {\em nolens volens}, at least a small tuning in
the weak scale.

An unfortunate consequence of this is that, even if the electroweak
scale is mostly natural, we may struggle to probe the associated
dynamics at the LHC. At best, we might hope that one or two new states
are anomalously light, such that we can either produce them on-shell,
or see their effects indirectly in rare processes.\footnote{Such light states might
be present for a variety of reasons. For example, they might be desirable
because they reduce fine-tuning (such as a light top squark and gluino in
SUSY), or they might arise because of symmetries (such as additional
pseudo-Goldstone boson states in composite Higgs models \cite{Gripaios:2009pe,Mrazek:2011iu,Barnard:2014tla}).} It is clear that discovery of such
states
will require painstaking work, including careful
scrutiny of all discrepancies between the data and SM predictions. 

In this work, we ask, in this vein, whether anomalies recently observed
in semileptonic decays of $B$-mesons at LHCb \cite{Aaij:2013iag,Aaij:2013qta,Aaij:2014ora} can be explained by a model in which a scale of a few TeV arises naturally
via strong dynamics. The necessary residual fine tuning required to
generate the electroweak scale can be achieved by making the Higgs
boson a pseudo-Goldstone boson (PGB) of global symmetries of the strong
dynamics sector \cite{Kaplan:1983fs,Georgi:1984af,Dugan:1984hq,Agashe:2004rs}. The Higgs potential (and thus the electroweak scale)
arises due to the breaking of the global symmetries by the SM gauging
and by couplings to fermions, and one can hope that there is an accidental
cancellation in the various contributions, whence a somewhat lower
electroweak scale emerges. The Yukawa couplings of the SM are assumed
to arise via the mechanism of partial compositeness \cite{Kaplan:1991dc}, which not
only provides a rationale for the structure of masses and mixings
observed in the quark sector, but also provides a paradigm for
suppressing large flavour-violating effects in processes involving the light fermions,
where the experimental constraints are strongest. 

The general framework of partial
compositeness is an obvious choice for
explaining the
anomalies,
which appear in processes involving second and third generation quarks,
and which appear to require new physics in muonic, but not electronic processes.
To fit the detailed structure of the anomalies, we hypothesize that they are due to the
presence in such a model of an anomalously light ({\em c.\  }TeV, as it turns out)
leptoquark. As pointed out in
\cite{Gripaios:2009dq},\footnote{In fact, ref.\ \cite{Gripaios:2009dq}, argued that
evidence for such leptoquarks should first appear in $b \rightarrow s
\mu \mu$ processes, precisely where the anomalies are now observed.} partial
compositeness models necessarily feature
a plethora of composite coloured fermion states, namely the composite quarks, and so
it would be something of a surprise if they did not also feature
composite
coloured scalar states, which could couple as leptoquarks or diquarks
\cite{Giudice:2011ak}. Moreover, one can easily arrange for a leptoquark state
to be rather lighter than the other resonances of the strong sector,
by making it a PGB of the same symmetry breaking that gives rise to the Higgs
boson.\footnote{The leptoquark is nevertheless expected to be
somewhat heavier than the Higgs \cite{Gripaios:2009dq}, both because it receives contributions to
its potential from the QCD coupling and because we expect that the
Higgs mass has been slightly tuned.}

A disadvantage of such models is that, being strongly coupled, we
cannot calculate {\em ad libitum}.
But we can use na\"{\i}ve dimensional analysis (NDA) to compute and make predictions
modulo $O(1)$ corrections. Using this framework, we find that the
anomalies single out one among the possible SM irreps that allow
leptoquark couplings, {\em viz.\ }a triplet under both $SU(3)_c$ and
$SU(2)_L$. This leptoquark is one of those identified in a recent
analysis~\cite{Hiller:2014yaa} of LHCb $B$ meson anomalies, in which
just two, non-vanishing leptoquark couplings (to $b$ quarks, $s$ quarks
and muons)
were invoked in an {\em ad hoc} fashion to fit the anomalies. 
In contrast, the model considered here is underpinned by a complete
(albeit presently uncalculable) framework for flavour physics, and all leptoquark
couplings are non-vanishing, with magnitude fixed by the degrees of compositeness of 
each of the SM fermion multiplets, giving 15 mixing parameters. In the quark
sector, all but
one of these parameters is fixed by measurements of quark masses and
the CKM matrix; there is more ambiguity in the lepton sector, but we
find that everything can be fixed by assuming that the mixings of the left
and right-handed lepton multiplets are comparable. This assumption is
a plausible one, from the point of view of the UV flavour dynamics, and
has the additional benefit that new physics (NP) corrections to the
most severely constrained flavour-violating observable, $\mu
\rightarrow e \gamma$, are minimized. 
As a result, we are
left with just 3 free parameters in the model: the mass, $M$, of the leptoquark, the
coupling strength, $g_\rho$, of the strong sector resonances, and the
degree of
compositeness, $\epsilon_3^q$, of the third generation quark doublet. 
Furthermore, all processes to which the leptoquark contributes (with the exception of meson mixing)
result in constraints on the single combination
$x \equiv \sqrt{g_\rho } \epsilon_3^q /M$. Thus the model is extremely
predictive. We find that the preferred range of $x$ corresponds to
plausible values of the 3 underlying parameters of the strongly
coupled theory (in which the weak scale is slightly tuned), namely $g_\rho \sim 4\pi$, $M \sim$ TeV, and
$\epsilon_3^q \sim 1$. Thus, $g_\rho$ and $\epsilon_3^q$ lie close to
their maximal values, meaning that one cannot evade future direct
searches at the LHC by scaling up $M$ and $g_\rho$. 

As for the existing bounds, we find that there is no obvious conflict,
but that there is potential to see effects in $\mu \to e \gamma$, $K^+
\rightarrow \pi^+ \nu \nu$, and $B^+
\rightarrow \pi^+ \mu^+ \mu^-$, in the near
future. Moreover, the required mass range for the leptoquark is not
far above that already excluded by LHC8, and so there is plenty of scope for
discovery in direct production at LHC13.

The outline is as follows. In the next Section, we describe the
data anomalies and review fits thereto using higher-dimensional SM
operators. We also show that they can be described by a leptoquark
carrying the representation $(\mathbf{\overline{3}},\mathbf{3},\frac{1}{3})$ of the $SU(3)\times
SU(2)\times U(1)$ gauge group. In \S \ref{sec:pc} we review the partial
compositeness and strong dynamics paradigms. We show how the leptoquark can accompany the Higgs as a PGB
of strong dynamics and exhibit symmetries that prevent proton decay,
{\em \&c.} In \S \ref{pheno}, we discuss important constraints on the model and describe the prospects for direct searches for
the leptoquark at LHC13 and indirect searches using flavour physics.

\section{Status of $b \to s \ell \ell$ fits and leptoquark quantum numbers} 
\label{phenobs}
The anomalies that we wish to explain were observed at LHCb in semileptonic $B$ meson decays involving a $b \to s$ quark transition.
These may be described via the low-energy, effective hamiltonian
\begin{equation}
{\mathcal H}_{\rm eff} = - 
\frac{4 G_F}{\sqrt{2}}\, (V_{ts}^\ast V_{tb}) \,  
\sum_{i}^{} C^{\ell}_i (\mu) \, {\mathcal O}^{\ell}_i (\mu) \, \, ,
\label{effH}
\end{equation}
where $\mathcal{O}^{\ell}_i$ are a basis of $\rm SU(3)_C \times U(1)_Q$-invariant 
dimension-six operators giving rise to the flavour-changing transition. The superscript $\ell$ denotes the lepton flavour in the 
final state $(\ell \in \{e,\mu,\tau\})$, and the operators
$\mathcal{O}^{\ell}_i$ are given in a standard basis by
\begin{eqnarray}
{\mathcal O}_7^{(')} &=& \frac{e}{16\pi^2} \,  
m_b \left (\bar s \sigma_{\alpha \beta} P_{R(L)} b \right )F^{\alpha \beta} \; , \nonumber
\\ 
{\mathcal O}^{\ell (')}_9 &=& \frac{\alpha_{\rm em}}{4 \pi} \,  
\left (\bar s \gamma_\alpha P_{L(R)} b \right ) (\bar \ell \gamma^\alpha \ell) 
\; , \; \rm \label{WilsonOps} \\ 
{\mathcal O}^{\ell (')}_{10} &=& \frac{\alpha_{\rm em}}{4 \pi} \, 
\left (\bar s \gamma_\alpha P_{L(R)} b \right ) 
(\bar \ell \gamma^\alpha \gamma_5 \ell). \nonumber
\end{eqnarray}
We neglect possible (pseudo-)scalar and tensor 
operators, since these have been
shown~\cite{Alonso:2014csa,Hiller:2014yaa} to be constrained to be too
small (in the absence of fine-tuning in the electron sector) to
explain LHCb anomalies. In the SM, the operator coefficients are lepton universal and the operators that have non-negligible coefficients are ${\mathcal O}_7$, ${\mathcal O}^{\ell}_9$, and ${\mathcal O}^{\ell}_{10}$, with
\begin{eqnarray}
C_7^{SM}&=&-0.319, \nonumber \\
C_9^{SM}&=&4.23, \label{SMWCs} \\
C_{10}^{SM}&=&-4.41. \nonumber
\end{eqnarray}
at the scale $m_b$~\cite{Khodjamirian:2010vf}. 

The first tension with the SM was observed last year in angular observables in the semileptonic decay $B \to K^* \mu^+ \mu^-$~\cite{Aaij:2013iag,Aaij:2013qta}.
The r\^{o}le of theoretical hadronic uncertainties in the discrepancy
is not yet clear~\cite{Khodjamirian:2010vf, Khodjamirian:2012rm,Jager:2012uw,Beaujean:2013soa,Lyon:2014hpa,Descotes-Genon:2014uoa,Altmannshofer:2014rta,Jager:2014rwa}. Nevertheless, several model-independent analyses~\cite{Descotes-Genon:2013wba,Altmannshofer:2013foa,Beaujean:2013soa,
Hurth:2013ssa,Horgan:2013pva}
have been performed on the $B \to K^* \mu^+ \mu^-$ decay data, as well
as on other, relevant, semileptonic and leptonic processes, allowing
for the possibility of new physics contributions to the effective
operators in eq. (\ref{WilsonOps}). 
There seems to be a consensus that, if only a single Wilson
coefficient is allowed to be
non-vanishing, then NP contributions to
the effective operator $\mathcal{O}_9^{\mu}$ are preferred,
with the NP coefficient $C_9^{NP}$ of this operator being
negative. A number of models of NP were proposed to explain this effect \cite{Gauld:2013qja,
 Datta:2013kja,Buras:2013dea,
 Buras:2013qja,Gauld:2013qba,
 Altmannshofer:2014cfa}.

Earlier this year LHCb measured another discrepancy in $B$ decays. To
wit, it was found that a certain ratio, $R_K$, of branching ratios of $B \to K \mu^+ \mu^-$ to $B \to K e^+ e^-$ 
lay 2.6$\sigma$ below the SM
prediction~\cite{Aaij:2014ora}. Specifically, the observable is defined as
\begin{equation}
R_K= \frac{
\int_1^6 dq^2 \frac{d \Gamma(B^+ \to K^+ \mu^+ \mu^-) \ }{dq^2}
}
{
\int_1^6 dq^2 \frac{d \Gamma(B^+ \to K^+ e^+ e^-)}{dq^2}
},
\label{RKeqn}
\end{equation}
where $q^2$ is the invariant mass of the di-lepton pair and the
integral is performed over the interval $q^2 \in [1,6] \textrm{ GeV}^2$.
Like the $B \to K^* \mu^+ \mu^-$ decay, these processes proceed via a
$b \to s \ell \ell$ transition. The observable
$R_K$ has the advantage of being theoretically well-understood, predicted to be almost
exactly 1 in the SM~\cite{Bobeth:2007dw} (specifically, $1.0003 \pm
0.0001$ when mass effects are taken into
account~\cite{Hiller:2003js}). A discrepancy in $R_K$ cannot be
explained by lepton-flavour-universal NP, nor by any of the sources of
theoretical uncertainty that might underlie the $B \to K^* \mu^+ \mu^-$
anomalies. Analyses and fits including the $R_K$ data and other
recent measurements were performed in
\cite{Hiller:2014yaa,Ghosh:2014awa,Hurth:2014vma,Altmannshofer:2014rta}.
Due to the lepton non-universality required by the $R_K$ data, these
analyses allowed the electronic and muonic Wilson coefficients to differ. They found that a negative contribution to $C_9^{\mu}$ remains favoured,
while contributions to electronic Wilson coefficients $C^e_i$ were found to be
consistent with zero, but could have large deviations therefrom, due to
larger experimental uncertainties in electronic measurements.

One could argue that the `axial-vector' basis, whilst
convenient for studying physics below the weak scale, is not
the most
natural choice in the context of models of NP above the weak scale,
which must respect the chiral gauge symmetries of the SM.  
In the absence of multiple couplings or particles that have been
somehow tuned (perhaps by additional symmetries), NP is likely to
generate operators that are coupled to a specific lepton chirality,
and thus aligned with a `chiral basis' 
in which $C_9=-C_{10}$, $C_9=C_{10}$, $C_9'=-C_{10}'$, $C_9'=C_{10}'$. Given this, the recent analyses have also made use of this basis~\cite{Hiller:2014yaa,Ghosh:2014awa,Hurth:2014vma,Altmannshofer:2014rta}
. They find that, when looking at NP contributions in a single Wilson coefficient at a time, the best fit in this basis is achieved by a negative contribution to $C_9^{\mu}=-C_{10}^{\mu}$.

Therefore, of the possible scalar leptoquarks,\footnote{For a review see \cite{Davidson:1993qk}.} which always generate contributions to one Wilson coefficient in the chiral basis, the obvious choice to explain the anomalies appears to be that with quantum numbers $(\mathbf{\overline{3}},\mathbf{3},1/3)$, which contributes to the combination $C_9^{\mu}=-C_{10}^{\mu}$ at tree level.\footnote{Vector leptoquarks with charges $(\mathbf{3},\mathbf{1},2/3)$ or $(\mathbf{3},\mathbf{3},2/3)$ also generate the required structure, but cannot be directly realized as Goldstone bosons.}
This leptoquark was already considered to explain $R_K$ in
\cite{Hiller:2014yaa}, in a scenario in which its only non-zero
couplings were to $b \mu$ and to $s \mu$.\footnote{Another interpretation of both the $R_K$ anomaly and a deviation seen at CMS, in the context of $R$-parity violating supersymmetry, was given in~\cite{Biswas:2014gga}.}

\section{Details of the composite model}
\label{sec:pc}
\subsection{Flavour structure and leptoquark couplings}
With the required quantum numbers of the leptoquark in hand, we now
embed the leptoquark in a composite Higgs model.\footnote{For a recent review see \cite{Bellazzini:2014yua}.} We assume, then, the presence of a new strong sector and of an elementary sector.
The strong sector is characterised by a mass scale $m_{\rho}$ and by a single coupling among the resonances, which is denoted by $g_{\rho}$.
We expect the strong sector in isolation to have a global symmetry
$\mathcal{G}$ which is spontaneously broken by the strong dynamics to
a subgroup $\mathcal{H}$. 
The SM gauge interactions are introduced in the strong sector by gauging a subgroup of $\mathcal{H}$. We identify the Goldstone bosons coming from the breaking $\mathcal{G} / \mathcal{H}$ 
with the Higgs boson $H$ and the leptoquark state $\Pi$.
So as to avoid large contributions to other flavour observables,
  we seek a model in which the coset space contains only $H$ and $\Pi$.

We postulate that the SM fermion Yukawa couplings are generated via the paradigm of partial compositeness \cite{Kaplan:1991dc}. The basic assumption is that elementary states $f^{a}_i$ 
(where $a \in \{q,u,d,\ell,e\}$ and $i$ is the family index) couple linearly to fermionic operators $\overline{\mathcal{O}}^a_i$ of the strong sector. For example, the relevant lagrangian 
required to generate the masses of the up quarks is, schematically,
\begin{equation}
\mathcal{L} \supset g_{\rho} \epsilon^q  \overline{\mathcal{O}}^q q + g_{\rho} \epsilon^u \overline{\mathcal{O}}^u u 
+ m_{\rho} \left(\overline{\mathcal{O}}^q \mathcal{O}^q +\overline{\mathcal{O}}^u \mathcal{O}^u \right) 
+ g_{\rho} \overline{\mathcal{O}^q} H \mathcal{O}^u.
\end{equation}
After electroweak symmetry breaking (EWSB), the resulting light mass eigenstates correspond to the SM fields and are given by linear combinations of the form  
\begin{equation}
 f^a_{SM} = \cos \theta^a \, f^a + \sin \theta^a \, \mathcal{O}^a  ,
\end{equation}
with $\sin \theta^a = O \left( \epsilon^a \right)$. Thus, the parameters $\epsilon^a_i$ have a physical meaning: they measure the degree of compositeness of the SM fields. 
If $\epsilon^a_i \lesssim 1$,  we have that (at leading order in $\epsilon$) $f_{SM} \approx f$ and the projections of the composite operators onto the SM fields are given by $ \left( \mathcal{O}^a \right)_{SM} \sim \epsilon^a f_{SM}$.
In this way, projecting operators such as $g_{\rho}
\overline{\mathcal{O}}^q H \mathcal{O}^u$ along the SM components, we
can read off the strength of the Yukawa interactions. 
In particular, for the the up and down quarks, we have
\begin{equation}
\left( Y_u \right)_{ij} \sim g_{\rho} \epsilon^q_i \epsilon^u_j \, , \qquad \left( Y_d \right)_{ij} \sim g_{\rho} \epsilon^q_i \epsilon^d_j.
\end{equation}
Throughout this Section, we use the symbol $\sim$ to
mean a relation that holds up to an unknown $O(1)$ coefficient whose
value is fixed by the uncalculable strong sector dynamics.
With an appropriate choice of the values of $\epsilon^q_i$,
$\epsilon^u_i$, and $\epsilon^d_i$, it is possible to reproduce the
hierarchy of the quark masses and the mixing angles of the CKM matrix.
We find
\begin{eqnarray}
\label{phenorel}
g_{\rho} v \epsilon^q_i \epsilon^u_i \sim m^u_i, \, \qquad g_{\rho} v \epsilon^q_i \epsilon^d_i  \sim m^d_i \\
\nonumber
\frac{\epsilon^q_1}{\epsilon^q_2} \sim \lambda , \, \qquad \frac{\epsilon^q_2}{\epsilon^q_3} \sim \lambda^2 , \, \qquad \frac{\epsilon^q_1}{\epsilon^q_3} \sim \lambda^3 , \,
\end{eqnarray}
where $v$ is the Higgs VEV, $\lambda=0.23$ is the Cabibbo angle and
$m^u_i$ and $m^d_i$ are the masses of the up- and down-type quarks, respectively. 
In our framework, then, the Yukawa sector is described by 10
parameters $(g_{\rho},\epsilon^q_i,\epsilon^u_i,\epsilon^d_i)$. The
phenomenological relations (\ref{phenorel}) can be used to reduce the
number of free parameters that we can use to fit the anomalies. Indeed, there are 8 independent relations in (\ref{phenorel}) and we choose to parametrize everything in terms of $g_{\rho}$ and $\epsilon^q_3$.
In the lepton sector, there is more arbitrariness in the values of $\epsilon^{\ell}_i$ and $\epsilon^e_i$. This is due to the fact that there are several mechanisms that can be envisaged for introducing 
mass terms in the neutrino sector. In order to make progress, we shall assume the
left and right mixing parameters to be of the same order, $\epsilon^e_i
\approx
\epsilon^{\ell}_i$. This assumption about the unknown flavour
  dynamics at high scales is a plausible one, but it also has the
  phenomenological advantage that it mitigates constraints on NP
  coming from lepton flavour violating (LFV) observables, such as $\mu \rightarrow e\gamma$,
  which are the most problematic flavour-violating observables
    for partial compositeness models \cite{Davidson:2007si,Redi:2013pga}. Indeed, physics at the scale $m_{\rho}$ generates a contribution to the radiative LFV decays of the form 
$\Gamma(\ell^i \to \ell^j \gamma) \sim \left| \epsilon^{\ell}_i \epsilon^{e}_j \right|^2+ \left| \epsilon^{\ell}_j \epsilon^{e}_i \right|^2$. Considering the mass constraints 
$\epsilon^{\ell}_i \epsilon^{e}_i = \frac{m^e_i}{g_{\rho} v} \delta_{ij}$, it is easy to show that $\left| \epsilon^{\ell}_i \epsilon^{e}_j \right|^2+ \left| \epsilon^{\ell}_j \epsilon^{e}_i \right|^2$
is minimized when
\begin{equation}
\label{minLFV}
\frac{\epsilon^{\ell}_i}{\epsilon^{\ell}_j} \sim \frac{\epsilon^{e}_i}{\epsilon^{e}_j} \sim \sqrt{\frac{m^{e}_i}{m^e_j}}.
\end{equation}
Evidently, this condition is implied by (but does not imply) our assumption that the
left and right leptonic mixings are equal.

In this way, we are able to fix all parameters in the lepton sector in
terms of $g_\rho$, and so all the NP effects of the model are
parameterized by $M$, $g_\rho$, and $\epsilon_q^3$. The phenomenological inputs and the expressions of the various mixing
parameters are summarised in Figs. \ref{masses} and \ref{mixing}.
\begin{center}
\begingroup
\begin{figure}
\begin{center}
\begin{tabular}{c c}
\toprule
 Fermion & Mass \\
\midrule
$e$ & 0.487 MeV \\
$\mu$ & 103 MeV\\
$\tau$ & 1.78 GeV\\ 
$d$ & $2.50^{+1.08}_{-1.03}$ MeV\\
$s$ & $47^{+14}_{-13}$ MeV \\
$b$ & $2.43 \pm 0.08$ GeV\\
$u$ & $1.10^{+0.43}_{-0.37}$ MeV\\
$c$ & $0.53 \pm 0.07$ GeV\\
$t$ & $150.7\pm 3.4$ GeV\\
\bottomrule
\end{tabular}
\end{center}
\caption{Values of running fermion masses at the scale $\mu=$ 1 TeV \cite{Xing:2007fb}. \label{masses}}
\end{figure}
\endgroup
\end{center}
\begin{center}
\begingroup
\begin{figure}
\begin{center}
\begin{tabular}{c c}
\toprule
Mixing Parameter & Value \\
\midrule
$\epsilon^q_1 = \lambda^3 \epsilon^q_3 $ & $1.15 \times 10^{-2} \, \epsilon^q_3$  \\
$\epsilon^q_2 = \lambda^2 \epsilon^q_3$ & $5.11 \times 10^{-2} \, \epsilon^q_3$ \\
\hline
$\epsilon_1^u= \frac{m_u}{v g_{\rho}} \frac{1}{\lambda^3 \epsilon^q_3}$ & $5.48 \times 10^{-4} / ( g_{\rho} \epsilon^q_3 )$ \\
$\epsilon_2^u= \frac{m_c}{v g_{\rho}} \frac{1}{\lambda^2 \epsilon^q_3}$ & $5.96 \times 10^{-2} / ( g_{\rho} \epsilon^q_3 )$ \\
$\epsilon_3^u= \frac{m_t}{v g_{\rho}} \frac{1}{\epsilon^q_3}$ & 0.866/$( g_{\rho} \epsilon^q_3 )$ \\
\hline
$\epsilon_1^d= \frac{m_d}{v g_{\rho}} \frac{1}{\lambda^3 \epsilon^q_3}$ & $ 1.24 \times 10^{-3} /( g_{\rho} \epsilon^q_3 )$ \\
$\epsilon_2^d= \frac{m_s}{v g_{\rho}} \frac{1}{\lambda^2 \epsilon^q_3}$ & $ 5.29 \times 10^{-3} /( g_{\rho} \epsilon^q_3 )$ \\
$\epsilon_3^d= \frac{m_b}{v g_{\rho}} \frac{1}{\epsilon^q_3}$ & $ 1.40 \times 10^{-2} ( g_{\rho} \epsilon^q_3 )$ \\
\hline
$\epsilon_1^{\ell}= \epsilon_1^{e}=\left(\frac{m_e}{g_{\rho} v}\right)^{1/2}$ & $ 1.67\times 10^{-3} /g^{1/2}_{\rho}$ \\
$\epsilon_2^{\ell}= \epsilon_2^{e}=\left(\frac{m_{\mu}}{g_{\rho} v}\right)^{1/2}$ & $ 2.43 \times 10^{-2} /g^{1/2}_{\rho}$ \\
$\epsilon_3^{\ell}= \epsilon_3^{e}=\left(\frac{m_{\tau}}{g_{\rho} v}\right)^{1/2}$ & 0.101/$g^{1/2}_{\rho}$ \\
\bottomrule
\end{tabular}
\caption{Partial compositeness mixing parameters and values. \label{mixing}}
\end{center}
\end{figure}
\endgroup
\end{center}
We may now determine the leptoquark couplings, as follows. Similarly to \cite{Giudice:2007fh}, below the scale of the strongly-coupled resonances we can describe the low energy physics by an effective field theory (EFT) of the form
\begin{equation}
\mathcal{L}= \frac{m^4_{\rho}}{g^2_{\rho}} \mathcal{L}^{(0)} \left( \frac{g_{\rho} \epsilon_i^a f^a_i}{m_{\rho}^{3/2}}, \frac{D^{\mu}}{m_{\rho}}, \frac{g_{\rho} H}{m_{\rho}}, \frac{g_{\rho} \Pi}{m_{\rho}} \right).
\end{equation}
In the strongly-coupled, UV theory we expect the presence of an
operator of the form  $g_{\rho} \Pi \overline{\mathcal{O}}^L
\mathcal{O}^Q$, where $\mathcal{O}^Q $ (or $\mathcal{O}^L$) is a composite operator with the same 
quantum numbers as a SM quark (or lepton). Below the scale $m_{\rho}$, this operator generates a contribution to $\mathcal{L}$ of the form $\sim g_{\rho} \epsilon^{\ell}_i \epsilon^q_j \Pi \ell_i q_j $.
At low energies, the renormalizable lagrangian of the model is
\begin{equation}
 \mathcal{L} =  \mathcal{L}_{SM} + \left( D^{\mu} \Pi \right) ^{\dagger} D_{\mu} \Pi - M^2 \Pi^{\dagger} \Pi +  \lambda_{ij} \, \overline{q}^c_{Lj} i \tau_2 \tau_a \ell_{Li} \, \Pi + \textrm{ h.c.},
\end{equation}
with $\lambda_{ij}=g_{\rho} c_{ij} \epsilon^{\ell}_i
\epsilon^q_j$, where we have omitted quartic terms involving $H$
and $\Pi$ that are not relevant to our discussion. Note that we have
explicitly re-introduced the $c_{ij}$ parameters that are expected to
be of $O(1)$, but are otherwise unknown. 
We summarise the values of the leptoquark couplings in Fig. \ref{couplings}.

\begin{center}
\begingroup
\begin{figure}
\begin{center}
\begin{tabular}{c c c c}
\toprule
$\lambda_{ij} / (c_{ij} g_{\rho}^{1/2} \epsilon_3^q )$ & $j=1$ & $j=2$ & $j=3$ \\
\midrule
$i=1$ & $1.92 \times 10^{-5}$ &  $8.53 \times 10^{-5}$ & $1.67 \times 10^{-3}$ \\
$i=2$ & $2.80 \times 10^{-4}$ &  $1.24 \times 10^{-3}$ & $2.43 \times 10^{-2}$ \\
$i=3$ & $1.16 \times 10^{-3}$ &  $5.16 \times 10^{-3}$ & $0.101$ \\
\bottomrule
\end{tabular}
\caption{Values of leptoquark couplings, $\lambda_{ij}$, where $i$ denotes the lepton generation label and $j$ the quark generation label. \label{couplings}}
\end{center}
\end{figure}
\endgroup
\end{center}

\subsection{Coset structure}
Here we supply a coset space construction that gives rise to the required SM quantum numbers for the Higgs and leptoquark fields.
First we describe the pattern of spontaneous breaking of the symmetry
of the strong sector $\mathcal{G/H}$, and the embedding of the SM
gauge group  
$SU(3)_C \times SU(2)_L \times U(1)_Y$ therein. We then discuss
additional symmetry structure required to avoid constraints from
nucleon decay and neutron-antineutron oscillations.

To build a coset, we start from the minimal composite Higgs model \cite{Agashe:2004rs}, in
which a single SM Higgs doublet arises 
 from the spontaneous breaking of $SO(5)$ to $SU(2)_H \times
 SU(2)_R$, with $H$ transforming as a $(\mathbf{2},\mathbf{2})$ of the unbroken subgroup.
We must now enlarge the coset space somehow to include the leptoquark
$\Pi$ and its conjugate $\Pi^{\dagger}$. To see how this may be
achieved, consider first a model with just the leptoquark and no Higgs
boson. This can be achieved using 
$SO(9)$ broken to $SU(4) \times SU(2)_{\Pi}$. The 6 Goldstone bosons,
$(\Pi ,\Pi^{\dagger})$, transform as
$(\mathcal{\mathbf{6}},\mathbf{3})$. 

Now form the direct product of $SO(5)$ and $SO(9)$ and consider the
coset space
\begin{equation}
 \label{coset}
 \frac{SO(9)\times SO(5)}{SU(4) \times SU(2)_{\Pi} \times SU(2)_H \times SU(2)_R}.
\end{equation}
This has, of course, the same Goldstone boson content as the two models above.
The trick is to somehow embed the SM gauge group in $\mathcal{H}$ so
as to get the right charges for $H$ and $\Pi$. To do so
we embed $SU(3)_C$ into $SU(4)$. Explicitly, $SU(4)$ contains a
maximal subgroup $SU(3)_C \times U(1)_{\psi}$, and the decomposition
of the 6-d irrep of $SU(4)$ under this group gives 
$\mathbf{6} = \mathbf{3}_{2/3} + \mathbf{\overline{3}}_{-2/3}$. 
We then embed $SU(2)_L$ as the diagonal subgroup of $SU(2)_{H} \times
SU(2)_{\Pi}$, while the hypercharge gauge group $U(1)_Y$ is embedded as
$T_Y= -\frac{1}{2} T_{\psi} + T_{3R} +T_X$, where $T_{\psi}$ generates $U(1)_{\psi}$, $T_{3R}$ belongs to the $SU(2)_R$ algebra, and $U(1)_X$ is an additional symmetry (under which the Higgs and the leptoquark are uncharged) which may be required to reproduce the correct SM hypercharge assignments.
It is now straightforward to show that the SM quantum numbers of $H$ and $\Pi$+$\Pi^{\dagger}$ are respectively $(\mathbf{1},\mathbf{2},1/2)$ and 
$(\mathbf{\overline{3}},\mathbf{3},1/3)$+$(\mathbf{3},\mathbf{3},-1/3)$,
as required.

We next need to show that the necessary Yukawa and leptoquark
couplings can be generated by linear mixing of the elementary fermions
of the SM with resonances of the strong sector carrying suitable
representations of the group $\mathcal{H}$. In fact, a number of
representations are available. One suitable assignment is
\begin{gather}
\mathcal{O}_q \sim (4,1,2,2)_{+1/2},~ \mathcal{O}_u \sim (4,1,1,1)_{+1/2},~ \mathcal{O}_d \sim (4,1,1,3)_{+1/2}, \\ \nonumber ~ \mathcal{O}_L \sim (4,3,2,1)_{-1/2},~ \mathcal{O}_e \sim (4,3,3,1)_{-1/2}
\end{gather}  
where the subscript denotes the charge under the $U(1)_X$ symmetry. It is straightforward to check that this assignment permits tri-linear couplings between the fermionic resonances and $H$ and $\Pi$ that yield the desired Yukawa and leptoquark couplings after mixing with the elementary fermions.

An advantage of this assignment is that we can use it to protect
$\Gamma (Z \rightarrow b\overline{b})$. This is
desirable since, with $g_\rho \sim 4 \pi$, there would otherwise be
sizable corrections to $\Gamma (Z \rightarrow b\overline{b})$, even
with $m_\rho \sim 10$ TeV. The protection cannot be achieved in exactly the
same way as in \cite{Agashe:2006at}, because the semi-direct product $(SU(2)_L \times SU(2)_R) \rtimes \mathbb{Z}_2$ used there is not a subgroup of $\mathcal{G}$. But we can instead use the symmetry $(SU(2)_H \times SU(2)_R) \rtimes \mathbb{Z}_2$, with much the same result.
In a nutshell (for more details,
see \cite{Gripaios:2014pqa}), the group\footnote{Or rather, strictly
speaking, its universal cover $Sp(2)$.} $SO(5) \subset G$ contains
not just $SU(2)_H \times SU(2)_R$, but also the larger subgroup
$\mathcal{K} \equiv (SU(2)_H \times SU(2)_R) \rtimes \mathbb{Z}_2$. We
require: (i) that this larger group be contained in $\mathcal{H}$;
(ii) that $b_L$ couple to a resonance of the strong sector
transforming as a $(2,2)$ under $SU(2)_H \times SU(2)_R \subset
\mathcal{K}$ and as either the trivial irrep or the sign irrep under
$\mathbb{Z}_2 \subset \mathcal{K}$; and (iii) that the coupling of
$b_L$ to the strong sector respect the subgroup $(U(1)_H \times
U(1)_R) \rtimes \mathbb{Z}_2 \subset \mathcal{K}$. With these three
requirements, a straightforward modification of the arguments given in
\cite{Agashe:2006at,Gripaios:2014pqa} shows that there can be no
corrections to $\Gamma (Z \rightarrow b\overline{b})$.

There is, however, a disadvantage with this assignment, in that the linear mixing between $\mathcal{O}_q$ and $q$ breaks the $SU(2)_L \times SU(2)_R$ custodial symmetry, which is often invoked to protect $\frac{m_W}{m_Z}$. Since $\epsilon^q_3 = 1$, these corrections are unsuppressed. 
Happily, we find thanks to $m_\rho \sim 10$ TeV and to the presence of light custodians \cite{Contino:2006qr,Pomarol:2008bh}, we are consistent with the bounds coming from EWPT observables.\footnote{Note that with the alternative assignment $\mathcal{O}_q \sim (4,2,1,1),~ \mathcal{O}_u \sim (4,2,2,2),~\mathcal{O}_d \sim (4,2,2,2),~\mathcal{O}_L \sim (4,2,1,1),~\mathcal{O}_e \sim (4,2,2,2)$, (for which an additional $U(1)_X$ is not necessary), the linear mixing between $\mathcal{O}_q$ and $q$ is $SU(2)_L \times SU(2)_R$ invariant, and corrections to $\frac{m_W}{m_Z}$ are suppressed by powers of $(\epsilon^u_3)^4 \ll 1$. But then one must relinquish custodial protection of $\Gamma (Z \rightarrow b\overline{b})$.}

The global symmetry $\mathcal{G}$ is broken explicitly by the gauging of the SM group, as well as by the linear couplings between the elementary and composite sector. As a result of
these breakings, the PGBs get a mass term. NDA suggests that the main
contribution to the effective potential of the Higgs comes from from the top Yukawa coupling. 
This implies a negative contribution to the Higgs mass parameter \cite{Agashe:2004rs}, which can trigger EWSB, and the resulting Higgs mass is expected to be of order $m^2_H \sim \frac{y^2_t}{16 \pi^2} m^2_{\rho}$. 
In contrast to the Higgs boson, the composite leptoquark gets its
dominant mass term contribution from QCD. The resulting leptoquark 
mass is of order $m^2_{\Pi} \sim \frac{g^2_s}{16 \pi^2} m^2_{\rho}$
and is positive-definite, avoiding the danger of
colour- and charge-breaking vacua.

We now move on to discuss constraints from nucleon decay, {\em \&c}.
In models with TeV scale strong dynamics, we cannot expect the accidentally symmetries of the SM that lead to conservation of baryon and lepton number to be preserved. This problem is exacerbated in our model with a light leptoquark state, since the SM gauge symmetry allows a $(\mathbf{\overline{3}},\mathbf{3},\frac{1}{3})$ leptoquark to couple to both $qq$ and $q \ell$, and thus mediate proton decay. 

We now assess whether additional global symmetries can be imposed to
prevent such decays. Our objective is to allow the coupling to $q
\ell$, but not that to $qq$. Evidently, then, $q$ and $\ell$ must
carry different charges, $e_q$ and $e_\ell$, say, under such a
symmetry.\footnote{We assume that all particles come in 1-d representations of the symmetry, so as not to have to introduce additional states.} We must now decide whether the leptoquark itself should carry charge or not. 

The easiest option to realise is for the leptoquark not to carry a
charge. Then the corresponding symmetry can lie outside of the $SO(9)$
group of which the $\Pi$ is a Goldstone boson. Then the leptoquark
coupling is allowed if $e_q + e_\ell = 0$. A problem with any such
symmetry is that it cannot forbid decays of 3 quarks to 3
anti-leptons. So, while the usual suspects, like $p \rightarrow e^+
\pi^0$ are forbidden, decays such as $p \rightarrow e^+ 2\overline{\nu}$ and $n \rightarrow 3 \overline{\nu}$ are not. 
In our framework the most stringent bound comes from searches for $p p
\rightarrow \mu^+ 2\overline{\nu}$ decays, where $\Gamma <  1.0 \times 10^{-63} \textrm{ GeV}$ \cite{Agashe:2014kda}.
The leading contribution to this processes is generated by the
dimension-9 operator $(qq d^{c\dagger})(\ell \ell e^{c\dagger})$, with
$\tau$ neutrinos. 
A NDA estimate gives
\begin{equation}
\Gamma (p \to \mu^+ \overline{\nu}_{\tau} \overline{\nu}_{\tau})_{NDA} = \frac{m^{11}_p}{(4 \pi)^3} 
\left( g^4_{\rho} \frac{\epsilon^d_1 (\epsilon^q_1)^2 \epsilon^{\ell}_2 (\epsilon^{\ell}_3)^2}{M^5} \right)^2 = 4.7 \times 10^{-53} \textrm{ GeV}^{-1}.
\end{equation}
It is then clear that the searches for such decays suffice to rule out
a model with compositeness at multi-TeV scales. 
In comparing with the bound, we have used the values 
\begin{gather}
\label{nominal}
M= 1\, \mathrm{TeV},~g_\rho =
  4\pi,~\mathrm{and}~\epsilon_3^q = 1,
\end{gather} 
and we shall continue to do so henceforth.

We need, therefore, to explore the alternative option, which is to
look for a symmetry that lies (at least partly) within $SO(9)$, such
that the leptoquark is charged. A simple expedient is to use the
$\mathbb{Z}_2 \subset SO(9)$ symmetry whose non-trivial element in the
defining representation of $SO(9)$ is the matrix $\begin{pmatrix} -I_6
  & 0 \\ 0 & I_3 \end{pmatrix}$, where $I_n$ is the $n \times n$
identity matrix. This element commutes with $SO(6) \times SO(3)$ (and
therefore is unbroken by the gauging of the SM subgroup) but
anti-commutes with the broken generators in $SO(9)/ SO(6) \times
SO(3)$, meaning that the leptoquarks transform under $\mathbb{Z}_2$ as
$\Pi \rightarrow - \Pi$. Now, by insisting that the $\mathbb{Z}_2$ be
unbroken by the strong dynamics and the couplings to elementary
fermions, the diquark coupling $\Pi qq$ is forbidden. Provided,
moreover, that the elementary $q$ and $\ell$ are assigned opposite
charges, the leptoquark coupling $\Pi \ell q$ is allowed. Yukawa
couplings can be retained by assigning the elementary $(u^c, d^c)$ and $e^c$ to have the same charges as $q$ and $\ell$, respectively.

Such a symmetry (which may be thought of as either a baryon or lepton
parity) stabilizes nucleons completely, and so also solves potential
problems from generic operators generated by the heavier resonances of
the strong dynamics. Its drawback is that it cannot forbid
neutron-antineutron oscillations,\footnote{For a review see \cite{Phillips:2014fgb}.} for which there are again strong
experimental constraints.
There are two dimension 9 operators in the EFT that could give a contribution to this process, namely
$qqqq (d^c d^c)^{\dagger}$ and $u^c d^c d^c u^c d^c d^c$.
The low-energy effects of these operators are subject to large
hadronic uncertainties; we estimate a rough bound on the necessary
scale as $\Lambda \gtrsim 100$ TeV.

In our leptoquark model, we expect to generate the operator
\begin{equation}
\frac{g_{\rho}^4 (\epsilon^q_1)^4 (\epsilon^d_1)^2}{M^5}qqqq (d^c d^c)^{\dagger}.
\end{equation}
Using the nominal values in (\ref{nominal}) and matching with the previous expression, we find
$\Lambda = 188$ TeV. Given the high dimension of the operator, this scale comes with a large uncertainty, but it would seem that we are safe.

Finally, we remark that we could, of course, invoke both symmetries discussed above, in order to forbid both nucleon decay and oscillations absolutely.

\section{Phenomenological analysis}
\label{pheno}
At tree level, the effects of the leptoquark on flavour physics
observables can be studied using the effective lagrangian
\begin{eqnarray}
\mathcal{L}_{LQ}^{eff} &=& \sum _{ij \ell k}
\frac{\lambda_{ij}(\lambda_{\ell k})^*}{4 M^2} \left[ (\overline{q}_j
  \tau^a \gamma^{\mu} P_L q_k)(\overline{\ell}_i \tau^a \gamma_{\mu}
  P_L \ell_{\ell})+ 3(\overline{q}_j \gamma^{\mu} P_L
  q_k)(\overline{\ell}_i \gamma_{\mu} P_L \ell_{\ell}) \right] ,
\label{effectiveL}
\end{eqnarray}
where $i,j, k,\ell \in \{1,2,3\}$ are generation indices. We work in a
basis where the CKM matrix acts on the up sector such that $q_j$ is
the quark doublet, $q_j=\left( V_{CKM}^{\dagger jk} u^k_L,~d^j_L
\right)^T$, and $\ell_i$ is the lepton doublet,
$\ell_i=\left(\nu^i,~e^i_L \right)^T$. We assume that the mass
differences between the components of the leptoquark triplet are small
compared to the masses themselves, so that the components can be assumed to have a common mass, $M$.
Therefore we may write
\begin{eqnarray}
 \mathcal{L}_{LQ}^{eff} = \sum _{ij \ell k} \frac{\lambda_{ij}(\lambda_{\ell k})^*}{2 M^2} & \left[
2 \left(\overline{d}_L \gamma^{\mu} d_L \right)_{kj} \left(\overline{e}_L \gamma_{\mu} e_L \right)_{\ell i}  + 
2 \left(\overline{u}'_L \gamma^{\mu} u^{\prime}_L \right)_{kj} \left(\overline{\nu}_L \gamma_{\mu} \nu_L \right)_{\ell i}  \right. \nonumber \\
&
+\left(\overline{d}_L \gamma^{\mu} d_L \right)_{kj} \left(\overline{\nu}_L \gamma_{\mu} \nu_L \right)_{\ell i}   
+\left(\overline{u}^{\prime}_L \gamma^{\mu} u^{\prime}_L \right)_{kj} \left(\overline{e}_L \gamma_{\mu} e_L \right)_{\ell i} \label{effLmass} \\
&
\left. +\left(\overline{u}^{\prime}_L \gamma^{\mu} d_L \right)_{kj} \left(\overline{e}_L \gamma_{\mu} \nu_L \right)_{\ell i} 
+\left(\overline{d}_L \gamma^{\mu} u^{\prime}_L \right)_{kj} \left(\overline{\nu}_L \gamma_{\mu} e_L \right)_{\ell i}  \right], \nonumber
\end{eqnarray}
where $u_L^{\prime j}=V_{CKM}^{\dagger jk} u_L^k$. All
unprimed fields are mass eigenstates.\footnote{We neglect neutrino
  masses.}
\bigskip

We now comment briefly on the qualitative consequences of the various operators
that appear above.
\begin{enumerate}[(i)]

\item Flavour changing neutral currents (FCNC) in the down quark sector 
 
 These are generated by the operators $\left(\overline{d}_L \gamma^{\mu} d_L \right)_{kj} \left(\overline{e}_L \gamma_{\mu} e_L \right)_{\ell i}$ and $\left(\overline{d}_L \gamma^{\mu} d_L \right)_{kj} \left(\overline{\nu}_L \gamma_{\mu} \nu_L \right)_{\ell i}$. They can mediate meson decays via the transitions $b \to s \ell \ell$, $b \to s \nu \nu$, $s \to d \ell \ell$, $s \to d \nu \nu$, $b \to d \ell \ell$ and $b \to d \nu \nu$. 
 
The $b \to s \ell \ell$ transition is the main motivation for this work and will be discussed in more detail below. 
The decays involving neutrinos can have large NP contributions,
because couplings to tau neutrinos are large in the partial
compositeness framework considered here. We provide a
quantitative analysis of the decays $B \rightarrow K^{(*)} \nu \nu$ and $K \to \pi \nu \nu$ below. 
Constraints on leptoquark couplings from measurements of
(lepton-flavour-conserving) $K$ and $B$ decays are summarized in
Fig.~\ref{BKconstraints} below, excluding $b \to s \ell \ell$ and $b
\to s \nu \nu$ processes, which will be discussed in more detail in the
text. Lepton-flavour-violating (LFV) processes, recently investigated in~\cite{Glashow:2014iga}, are also possible in our set-up, but current bounds on these are weak. We will comment more on LFV processes in \S~\ref{otherconstraints}.

\item FCNC in the up quark sector 

These are generated by the operators $\left(\overline{u}^{\prime}_L \gamma^{\mu} u^{\prime}_L \right)_{kj} \left(\overline{\nu}_L \gamma_{\mu} \nu_L \right)_{\ell i}$ and $\left(\overline{u}^{\prime}_L \gamma^{\mu} u^{\prime}_L \right)_{kj} \left(\overline{e}_L \gamma_{\mu} e_L \right)_{\ell i}$. They can mediate decays of charmed mesons via $c \to u \ell \ell$ and $c \to u \nu \nu$ transitions. 
Bounds on these processes are weak, and we know of no bounds for decays with $\tau$ leptons or neutrinos in the final state, which would receive the largest NP contributions. 
These operators can also generate top decays into $u$ or $c$ quarks
plus a pair of charged leptons or of neutrinos. The rates of these decays will be very small relative to current limits on FCNC top quark decays~\cite{Agashe:2014kda} (which in any case search specifically for $t \to Z q$, meaning they cannot be directly applied to leptoquarks).
Since current measurements of FCNC in the up sector do not provide
strong constraints on our model, we will not discuss them further.

\item Charged currents

These are generated by the operators $\left(\overline{u}^{\prime}_L \gamma^{\mu} d_L \right)_{kj} \left(\overline{e}_L \gamma_{\mu} \nu_L \right)_{\ell i}$ and $\left(\overline{d}_L \gamma^{\mu} u^{\prime}_L \right)_{kj} \left(\overline{\nu}_L \gamma_{\mu} e_L \right)_{\ell i}$.
Processes generated by these operators are also present at tree level
in the SM, so NP contributions are not expected to be large relative
to the SM predictions. The largest NP rates will occur in processes with $\tau$ or $\nu_{\tau}$ in the final state.
\end{enumerate}

With these considerations in mind, in the remainder of this Section we
discuss the values of the model parameters that are needed to fit recent $B$-decay anomalies and then list important constraints on the model and predictions for its effects in other processes.

\subsection{Anomalies in $B$ decays}
\label{anomalies}

\subsubsection{Fit to muonic $\Delta B = \Delta S =1$ processes}
We consider recent results of \cite{Altmannshofer:2014rta}, in which a fit
to all available data on muonic (or lepton-universal) $\Delta B =
\Delta S =1$ processes is described. 
A part of that work involved allowing one Wilson Coefficient (or chiral combination thereof) to vary while assuming all other coefficients are set to their SM values (for details of the fit please see \cite{Altmannshofer:2014rta}). The best fit value found in this way for the chiral combination relevant to our leptoquark is $C_9^{NP \mu}=-C_{10}^{NP \mu}=-0.55$, with 1$\sigma$ and 2$\sigma$ ranges
\begin{eqnarray}
C_9^{NP \mu}=-C_{10}^{NP \mu} \in \left[-0.74, -0.36 \right]~~~(\mathrm{at}~1\sigma), \\
C_9^{NP \mu}=-C_{10}^{NP \mu} \in \left[-0.95, -0.19 \right]~~~(\mathrm{at}~2\sigma).
\end{eqnarray}
It can be seen, by comparing the effective leptoquark lagrangian in (\ref{effLmass}) with the effective hamiltonian in (\ref{effH}), that, for our model,
\begin{equation}
C_9^{\mu NP}=-C_{10}^{\mu NP}=\left[ \frac{4 G_F e^2 (V^*_{ts}V_{tb})}{16 \sqrt{2} \pi^2} \right]^{-1} \frac{\lambda^*_{22} \lambda_{23}}{2 M^2}
= -0.24 \, c^*_{22} c_{23} (\epsilon^q_3)^2 \left( \frac{M}{\mathrm{TeV}} \right)^{-2} \left( \frac{g_{\rho}}{4 \pi} \right),
\end{equation}
and the requirements on the parameters are
\begin{eqnarray}
\mathrm{Re}(c^*_{22} c_{23}) = 2.30 \left( \frac{4 \pi}{g_{\rho}} \right) \left( \frac{1}{\epsilon_3^q} \right)^2 \left( \frac{M}{\mathrm{TeV}} \right)^2 ~~~(\mathrm{Best~fit}), \\
\mathrm{Re}(c^*_{22} c_{23}) \in \left[1.50, 3.08 \right] \left( \frac{4 \pi}{g_{\rho}} \right) \left( \frac{1}{\epsilon_3^q} \right)^2 \left( \frac{M}{\mathrm{TeV}} \right)^2 ~~~(\mathrm{at}~1\sigma), \\
\mathrm{Re}(c^*_{22} c_{23}) \in \left[0.79, 3.96 \right] \left( \frac{4 \pi}{g_{\rho}} \right) \left( \frac{1}{\epsilon_3^q} \right)^2 \left( \frac{M}{\mathrm{TeV}} \right)^2 ~~~(\mathrm{at}~2\sigma).
\end{eqnarray}.

Thus, if this anomaly is to be explained, there are 3 immediate implications for the parameters of our model:
\begin{enumerate}
 \item the mass of the leptoquark states should be low enough, $M \lesssim 1$ TeV, to be within the reach of the second run of the LHC;
 \item the left-handed doublet of the third quark family should be largely composite, $\epsilon_3^q \sim 1$;
 \item the composite sector must be genuinely strongly interacting, $g_{\rho} \sim 4 \pi$.
\end{enumerate}
Indeed, if any one of these does not hold then we are forced to set
$\mathrm{Re}(c^*_{22} c_{23}) \gg 1$, implying an inconsistency with the
EFT paradigm described in the previous Section.

\subsubsection{$R_K$}
$R_K$, as defined in eq. (\ref{RKeqn}), has been recently measured by
LHCb to be $R_K=0.745^{+0.090}_{-0.074} \pm
0.036$~\cite{Aaij:2014ora}. Roughly, adding errors in quadrature, we
therefore take the measured value at the 1$\sigma$ level to be within
the range $[0.66,0.84]$. The model prediction, including the effect of the leptoquark, is given approximately by 
\begin{equation}
\label{RKeqn2}
R_K \approx \dfrac{\left| C^{SM}_{10} + C^{\mu \mathrm{NP}}_{10} +C^{\mu'}_{10} 
\right|^2 +  \left| C^{SM}_9 +C^{\mu \mathrm{NP}}_{9} +C^{\mu'}_{9}  \right|^2}
{\left| C^{SM}_{10} + C^{e \mathrm{NP}}_{10} +C^{e'}_{10}  \right|^2 +  \left| 
C^{SM}_9 +C^{e \mathrm{NP}}_{9} +C^{e'}_{9}  \right|^2},
\end{equation}
which can be found from the full expression by neglecting the coefficient of the dipole operator, $C_7$. (In the SM $C_7$ has a magnitude less than 10\% that of $C_9$ or $C_{10}$, and NP contributions to it are constrained small by the measured branching ratio of $B \rightarrow X_s \gamma$). The NP values of the Wilson coefficients are
\begin{eqnarray}
C_9^{\mu NP} = -C_{10}^{\mu NP} &=& -0.24 \, c^*_{22} c_{23} (\epsilon^q_3)^2 \left( \frac{M}{\mathrm{TeV}} \right)^{-2} \left( \frac{g_{\rho}}{4 \pi} \right), \\
C_{9}^{e NP} = -C_{10}^{e NP} &=& - 1.2 \times 10^{-3} c^*_{11} c_{23} (\epsilon^q_3)^2 \left( \frac{M}{\mathrm{TeV}} \right)^{-2} \left( \frac{g_{\rho}}{4 \pi} \right).
\end{eqnarray}
The values of $C_9^{SM}$ and $C_{10}^{SM}$ are given in eq. (\ref{SMWCs}). 

We see that, due to the structure of partial compositeness, NP contributions in the decay $B^+ \to K^+ e^+ e^-$ are negligible. Neglecting these and the quadratic terms in $C_{9,10}^{\mu NP}$, 
we obtain
\begin{equation}
\mathrm{Re}(c^*_{22} c_{23}) \in \left[1.42,2.98 \right] \left( \frac{4 \pi}{g_{\rho}} \right) \left( \frac{1}{\epsilon_3^q} \right)^2 \left( \frac{M}{\mathrm{TeV}} \right)^2~~~(\mathrm{at}~1\sigma).
\end{equation}

The allowed region thus has reasonable overlap with the $1\sigma$ region found above using a fit to muonic $\Delta B=\Delta S=1$ observables. Therefore our model is able to fit the muonic data and $R_K$ with no tension between the two. 
Of course, this is hardly surprising as several
works~\cite{Alonso:2014csa,Ghosh:2014awa,Altmannshofer:2014rta} have
pointed out the compatibility of the $b \to s \mu \mu$ data with $R_K$
if the NP is predominantly in the muon sector, rather than the
electron sector. This feature is automatic in models with
partial compositeness.

\subsection{Important constraints and predictions}

The largest couplings of the composite leptoquark are to third
generation quarks and leptons. Therefore, generically, the most
important constraints and predictions will be in processes involving
third generation quarks and fermions in initial or final states and
also processes with third-generation fermions in a loop.\footnote{Of
  course, this is only true generically, since the sensitivity depends
not only on the size of the NP contribution, but also on the
experimental feasibility and also the size and nature of the competing
SM contributions.}
This Section will look at some of these processes, discussing implications of current measurements on our model, as well as highlighting promising channels for probing our scenario with future measurements.

\subsubsection{$b \rightarrow s \nu \nu$}

Due to the $SU(2)_L$ structure of the leptoquark, it will couple to
neutrinos as well as charged leptons and thus induce $b \to s \overline{\nu} \nu$ transitions. The importance of this channel in general for pinning down NP has been recently emphasised in \cite{Buras:2014fpa}. 
These $B \rightarrow K^* \nu \overline{\nu}$ and $B \rightarrow K \nu
\overline{\nu}$ decays are good channels to look for large effects
from the composite leptoquark we consider. Indeed,
since the identity of the neutrino cannot be determined in these
experiments, large contributions from the processes involving tau
neutrinos are expected in our model. Thus our model predicts a much
larger rate than that expected in models where NP couples only to the second generation lepton doublet.

Current NP bounds from these decays can be found in
\cite{Buras:2014fpa}, which are quoted in terms
of ratios to Standard Model predictions. With a slight alteration of
the notation of \cite{Buras:2014fpa}, so as not to cause confusion
with the notation used here, the relevant quantities, and the limits
thereon, are
\begin{equation}
R_K^{*\nu\nu} \equiv \frac{\mathcal{B} \left( B \rightarrow K^* \nu \overline{\nu}\right)}{\mathcal{B}\left({B \rightarrow K^* \nu \overline{\nu}}\right)_{SM}} < 3.7,
\end{equation}
and
\begin{equation}
R_K^{\nu\nu} \equiv \frac{\mathcal{B} \left( B \rightarrow K \nu \overline{\nu} \right) }{\mathcal{B} \left( B \rightarrow K \nu \overline{\nu} \right)_{SM} } < 4.0.
\end{equation}
The leptoquark can in principle induce transitions involving any combination of neutrino flavours, since it couples to all generations and also has flavour-violating couplings. There will be interference between NP and SM processes only in flavour-conserving transitions.
The NP contributions to the $\overline{\nu}_{\tau} \nu_{\tau}$  and $\overline{\nu}_{\mu} \nu_{\mu}$ processes will induce a shift from unity in $R_K^{\nu\nu}$ and $R^{(*)\nu\nu}_K$ given by
\begin{eqnarray}
\nonumber
\Delta(R^{(*)\nu\nu}_K)^{\tau \tau}  &=&  \left[ 0.220  \, \textrm{Re} (c^*_{32} c_{33})  + 0.0363 \left|c^*_{32} c_{33} \right|^2 (\epsilon^q_3)^2 \left( \frac{M}{\mathrm{TeV}} \right)^{-2} \left( \frac{g_{\rho}}{4 \pi} \right) \right] (\epsilon^q_3)^2 \left( \frac{M}{\mathrm{TeV}} \right)^{-2} \left( \frac{g_{\rho}}{4 \pi} \right), \\
\nonumber
\Delta(R^{(*)\nu\nu}_K)^{\mu \mu}  &\approx&  1.27 \times 10^{-2}  \, \textrm{Re} (c^*_{32} c_{33}) \,  (\epsilon^q_3)^2 \left( \frac{M}{\mathrm{TeV}} \right)^{-2} \left( \frac{g_{\rho}}{4 \pi} \right).
\end{eqnarray}
(The expression for $\Delta(R^{(*)}_K)^{\mu \mu}$ is approximate, because we have kept only the interference term with the Standard Model, which is large compared to the term from purely NP contributions.)
The next biggest contribution comes from $\overline{\nu}_{\mu}
\nu_{\tau}$ and  $\overline{\nu}_{\tau} \nu_{\mu}$ final states. In these cases, there is no interference with the SM and the contribution is 
\begin{equation}
\Delta(R^{(*)\nu\nu}_K)^{\mu \tau} + \Delta(R_K^{(*)\nu\nu})^{\tau \mu} =  2.10 \times 10^{-3} \left( |c^*_{22} c_{33}|^2 +|c^*_{32}c_{23}|^2 \right)   (\epsilon^q_3)^4 \left( \frac{M}{\mathrm{TeV}} \right)^{-4} \left( \frac{g_{\rho}}{4 \pi} \right)^2.
\end{equation}
As is clear from these equations, the most important contribution
comes from the $\overline{\nu}_{\tau} \nu_{\tau}$ process. It is
possible to pass the bound $\Delta(R^{(*)\nu\nu}_K)^{\tau \tau} <2.7$
in a large fraction of the parameter space. Furthermore, large
deviations in $R_K^{\nu\nu}$ and $R_K^{*\nu\nu}$ ($\sim 25\%$ of the SM contribution) represent an interesting prediction of our composite leptoquarks scenario, which will be testable at the upcoming Belle II experiment~\cite{Buras:2014fpa,Aushev:2010bq}.
Our prediction can be compared with the case in which the leptoquark has only muonic couplings, in which the contributions to $\Delta(R^{(*)\nu\nu}_K)$ are $\lesssim 5\%$ (see section 4.5 of \cite{Buras:2014fpa}).

\subsubsection{$K^+ \rightarrow \pi^+ \nu \nu$}
Given that measurements involving neutrinos have the ability to probe some of the largest couplings in our model -- those involving third generation leptons -- it is necessary to check other rare meson decays with final state neutrinos. 

Following~\cite{Hurth:2008jc}, (but rescaling the bound given there to match the
slightly more recent measurement in \cite{Agashe:2014kda}), the
measurement of $\mathcal{B}(K^+ \rightarrow \pi^+ \nu \nu)$ produces a
bound (at 95\% confidence level) on the real NP coefficient $\delta
C_{\nu \bar{\nu}}$ (defined in \cite{Hurth:2008jc}) of
\begin{equation}
\delta C_{\nu \bar{\nu}} \in [-6.3,2.3].
\end{equation}
The branching ratio is given in terms of $\delta C_{\nu \bar{\nu}}$ by
\begin{equation}
\mathcal{B}(K^+ \rightarrow \pi^+ \nu \nu) = 8.6(9) \times 10^{-11} [ 1+ 0.96\delta C_{\nu \bar{\nu}}+ 0.24 (\delta C_{\nu \bar{\nu}})^2].
\end{equation}
Our leptoquark contributes to $\delta C_{\nu \bar{\nu}}$ as
\begin{equation}
\delta C_{\nu \bar{\nu}} = 0.62~ \mathrm{Re} (c_{31} c^*_{32}) \left( \frac{g_{\rho}}{4 \pi} \right) \left( \epsilon_3^q \right)^2 \left( \frac{M}{\mathrm{TeV}}\right)^{-2},
\end{equation}
via the dominant process involving a pair of tau neutrinos. So with $c_{31} \sim c_{32} \sim O(1)$, and $M \sim$ TeV, our scenario passes current bounds.

However the NA62 experiment, due to begin data-taking in 2015, will measure $\mathcal{B}(K^+ \rightarrow \pi^+ \nu \nu)$ to an accuracy of 10\% of the SM prediction~\cite{Rinella:2014wfa}. This means it will be able to shrink the bounds on $\delta C_{\nu \bar{\nu}}$ to
\begin{equation}
\delta C_{\nu \bar{\nu}} \in [-0.2,0.2]
\end{equation}
at 95\%. 
Thus, if $c_{31} \sim c_{32} \sim O(1)$ and $M \sim$ TeV, measurements at NA62 will be sensitive to our leptoquark. 

\subsubsection{Meson mixing}
The leptoquark we consider can mediate mixing between neutral mesons via box diagrams. This effect will be largest in $B_s$ mesons. 
From \cite{Davidson:1993qk}, the bound produced on the leptoquark
couplings when both leptons exchanged in the box are taus (the
dominant contribution in our scenario) is
\begin{equation}
|\lambda_{33}\lambda_{32}^*|^2 < \frac{196\pi^2 M^2 \Delta m_{B_s^0}^{NP}}{f^2_{B_s^0}m_{B_s^0}}.
\end{equation}
From \cite{Lenz:2011ti}, $f_{B^0_s}=0.231$ GeV, and 
\begin{equation}
\Delta m_{B_s^0}^{SM} = (17.3 \pm 2.6) \times 10^{12} \hbar s^{-1} = (1.14 \pm 0.17) \times 10^{-8} \mathrm{ MeV},
\end{equation}
while from \cite{Agashe:2014kda}, the measured value of the mass splitting is
\begin{equation}
\Delta m_{B_s^0} = 17.69 \times 10^{12} \hbar s^{-1} = 1.2 \times 10^{-8} \mathrm{ MeV}.
\end{equation}
Taking the uncertainty in the prediction to be roughly the size of the NP contribution, $|\Delta m_{B_s^0}^{NP}/\Delta m_{B_s^0}^{SM}|<0.15$ (as in \cite{Hiller:2014yaa}), then
\begin{equation}
|\lambda_{33}\lambda_{32}^*|^2 < 0.017 \left( \frac{M}{\mathrm{TeV}} \right)^2.
\end{equation}
In terms of the parameters of our model this becomes
\begin{equation}
|c_{33}c_{23}^*| < 4.2 \left( \frac{4 \pi}{g_{\rho}} \right)^2 \left(\frac{M}{\mathrm{TeV}} \right)^2 \left( \frac{1}{\epsilon^q_3} \right)^4.
\end{equation}
We are able to pass this bound taking $O \left( 1 \right)$ values for $c_{33}$ and $c_{23}$ and taking the other parameters at values necessary to fit the anomalies as discussed above. The leptoquark will also contribute to mixing of other neutral mesons. However bounds from the measurement of mixing observables are generally weaker than bounds from meson decays (see eg.~\cite{Saha:2010vw}).

\subsubsection{$\mu \rightarrow e \gamma$ and other radiative processes}
The leptoquark has only left handed couplings, meaning that we will not get chiral enhancements to the branching ratio of $\mu \rightarrow e \gamma$. Nevertheless, the bound on $\mathcal{B}(\mu \rightarrow e \gamma)$ is tight enough to be relevant for the model.
The largest contributions come from diagrams with a loop
containing either a top or a bottom quark, together with the
leptoquark. The most recent measurement was performed by the MEG collaboration
\cite{Adam:2013mnn}, who found a bound at $90\%$ confidence level of
$\mathcal{B}(\mu^+ \rightarrow e^+ \gamma) < 5.7 \times 10^{-13}$.
Using the formula for the rate given in~\cite{Davidson:1993qk}, and neglecting all but the processes involving 3rd generation quarks in the loop,
\begin{equation}
\left| \lambda_{23}^* \lambda_{13} \right| < 7.3 \times 10^{-4} \left( \frac{M}{\mathrm{TeV}} \right)^2,
\end{equation}
which amounts to a bound on $c^*_{23}c_{13}$ of 
\begin{equation}
\left| c_{23}^*c_{13} \right| < 1.4 \left( \frac{4 \pi}{g_{\rho}} \right) \left(\frac{M}{\mathrm{TeV}} \right)^2 \left( \frac{1}{\epsilon^q_3} \right)^2.
\end{equation}
This turns out to be a strong constraint for our model. Given that our
EFT paradigm assumes $c_{ij} \sim O(1)$, the bound is, roughly,
saturated. 

Given our flavour structure we expect an even larger contribution to
$\tau \rightarrow \mu \gamma$ than to $\mu \rightarrow e
\gamma$. However the current bound on the branching ratio of this
process is $\mathcal{B}(\tau \rightarrow \mu \gamma)< 4.4 \times
10^{-8}$ \cite{Agashe:2014kda}, which is several orders of magnitude
larger than the model prediction. 

The process $b \to s \gamma$ can be generated via similar
diagrams.
Current bounds on this process, which leave room for NP contributions up to about $30\%$ of the SM prediction, lead to a bound on the combination $|c_{33}^*c_{32}|$ of roughly $|c_{33}^*c_{32}| \lesssim 100 \left( \frac{4 \pi}{g_{\rho}} \right) \left(\frac{M}{\mathrm{TeV}} \right)^2 \left( \frac{1}{\epsilon^q_3} \right)^2$.

\subsubsection{Comments on other constraints and predictions}
\label{otherconstraints}
Despite the fact that contributions from leptoquark diagrams will be largest for processes containing taus (or tau neutrinos) in the final state, we have not yet mentioned any bounds from meson decays with $\tau$ leptons in the final state. This is because existing bounds are very weak due to the relative difficulty of tau measurements. The current bound~\cite{Flood:2010zz} on the decay $B \to K \tau^+ \tau^-$ from BaBar, $\mathcal{B}(B \to K \tau^+ \tau^-) < 3.3 \times 10^{-3}$, is several orders of magnitude larger than the NP prediction. Likewise the recent Belle measurement of $\mathcal{B}(B^+ \to \tau^+ \nu)$~\cite{Abdesselam:2014hkd} has error bars much larger than the NP contribution (as does the SM prediction).

We have discussed $b \to s \ell \ell$ processes and anomalies in previous subsections. Bounds from meson decays mediated by other FCNC processes in the down sector are summarised in Fig.~\ref{BKconstraints}. The most constraining of these measurements is from the branching ratio of $B^+ \to \pi^+ \mu^+ \mu^-$, for which the bound is approximately saturated.

Our leptoquark can appear in diagrams which contribute to the muon
anomalous magnetic moment, an observable which currently has a $2.2
\mbox{--} 2.7 \sigma$ discrepancy with SM
calculations~\cite{Bennett:2006fi}. However, as was pointed out
in~\cite{Cheung:2001ip, Chakraverty:2001yg}, if a leptoquark couples
only to one chirality of muon, as is the case for us, the couplings
would need to be very large to explain the measurement. Our scenario
produces a prediction several orders of magnitude too small (for a
mass of $O$(1 TeV)), and so does nothing to alleviate the current
tension between the SM and experiment.

One hallmark of our model is that there should be only very small NP
effects in the electron sector. So decay measurements involving electrons should see no significant deviations from the Standard Model in our scenario.
A recent paper~\cite{Altmannshofer:2014rta} contains a table with predictions of ratios of observables with muons in the final state versus those with electrons for $b \rightarrow s \ell \ell$ processes. The predictions of our leptoquark model will, to a good approximation, coincide with those of the third column of their table, which contains the predictions for a scenario with NP only in $C_9^{\mu NP}=-C_{10}^{\mu NP}=-0.5$.

The leptoquark we consider will mediate lepton flavour
violating processes. However we find that all current bounds are well
above rates predicted for the leptoquark contribution. Lepton flavour
violation in the context of $B$ decays was recently discussed in
detail in~\cite{Glashow:2014iga}. There, the authors consider a model
in which, similarly to our case, the NP contributions to $b \to s \ell
\ell$ decays arise in a $V-A$ structure
(ie. $C_9^{\ell}=-C_{10}^{\ell}$) and the largest effects are in the
third generation of quarks and leptons. Interestingly, a special
  case of our model can be made to fit into their framework, if we take all the
$O(1)$ coefficients $c_{ij}$ to be equal (and for simplicity, equal to
1). Then the coupling denoted $G$ in~\cite{Glashow:2014iga} is given
by $G=(g_{\rho}/M^2)\, (\epsilon_3^q)^2 \, \frac{m_{\tau}}{v}$, and the mixing
matrices $U_{L3i}^{\ell}$ and $U_{L3i}^d$ therein are given by
$U^d_{L3}=(\lambda^3,\lambda^2,1)$ and
$U^{\ell}_{L3i}=\sqrt{m_i/m_{\tau}}$. With these choices, we find that
all bounds quoted in~\cite{Glashow:2014iga} are comfortably satisfied by the composite leptoquark model. More precise bounds on LFV processes will certainly provide an interesting test of our model and other lepton non-universal scenarios.

Another recent paper~\cite{Hiller:2014ula} proposes double ratios of branching ratios as clean probes of NP that is not lepton universal and couples to right-handed quarks. Since the leptoquark we consider has no couplings to right-handed quarks, measurements of these would be a useful test of the model if the $B$ anomalies persist.

\begin{center}
\begingroup
\begin{figure}
\begin{center}
\begin{tabular}{c c l l}
\toprule
Decay & (ij)(kl)$^*$ & $|\lambda_{ij} \lambda^*_{kl}| / \left( \frac{M}{\rm{TeV}} \right)^2 $ & $|c_{ij} c^*_{kl}| \left( \frac{g_{\rho}}{4 \pi} \right) \left( \epsilon_3^q \right)^2 / \left( \frac{M}{\rm{TeV}} \right)^2$ \\
\midrule
$K_S \to e^+ e^-$ & $ (12)(11)^*$ &  $< 1.0$ & $< 4.9 \times 10^7$ \\
$K_L \to e^+ e^- $ & $(12)(11)^*$ &  $< 2.7 \times 10^{-3}$ & $< 1.3 \times 10^{5}$ \\
$\dagger~K_S \to \mu^+ \mu^-$ & $(22)(21)^*$ &  $< 5.1 \times 10^{-3}$ & $< 1.2 \times 10^3$ \\
$K_L \to \mu^+ \mu^- $ & $(22)(21)^*$ &  $< 3.6 \times 10^{-5}$ & $< 8.3$ \\
$K^+ \to \pi^+ e^+ e^-$ & $(11)(12)^*$ &  $< 6.7 \times 10^{-4}$ & $< 3.3 \times 10^4$ \\
$K_L \to \pi^0 e^+ e^- $ & $(11)(12)^*$ &  $< 1.6 \times 10^{-4}$ & $< 7.8 \times 10^{3}$ \\
$K^+ \to \pi^+ \mu^+ \mu^-$ & $(21)(22)^*$ &  $< 5.3 \times 10^{-3}$ & $< 1.2 \times 10^3$ \\
$K_L \to \pi^0 \nu \bar{\nu} $ & $(31)(32)^*$ &  $< 3.2 \times 10^{-3}$ & $< 42.5$ \\
$\dagger~B_d \to \mu^+ \mu^-$ & $(21)(23)^*$ &  $< 3.9 \times 10^{-3}$ & $< 46.0$ \\
$B_d \to \tau^+ \tau^- $ & $(31)(33)^*$ &  $< 0.67$ & $< 4.6 \times 10^{2}$ \\
$\dagger~B^+ \to \pi^+ e^+ e^-$ & $(11)(13)^*$ &  $< 2.8 \times 10^{-4}$ & $< 6.9 \times 10^2$ \\
$\dagger~B^+ \to \pi^+ \mu^+ \mu^-$ & $(21)(23)^*$ &  $< 2.3 \times 10^{-4}$ & $< 2.7$ \\
\bottomrule
\end{tabular}
\caption{90\% confidence level bounds~\cite{Saha:2010vw} on leptoquark couplings from branching ratios of
  (semi-)leptonic meson decays involving $b \to d$ and $s \to d$,
  rescaled to $M= 1$ TeV. A dagger denotes bounds that have been
  rescaled to newer measurements~\cite{Agashe:2014kda}. The final
  column gives bounds on partial compositeness parameters in units of
  the nominal values in (\ref{nominal}).}
\label{BKconstraints}
\end{center}
\end{figure}
\endgroup
\end{center}

\subsection{Direct searches at the LHC}
\label{LHC}
If the leptoquark is light enough, as the arguments in
\S~\ref{anomalies} suggest it should be, it will be pair-produced at
the LHC with sizable cross-section via QCD interactions. The
leptoquark field comprises 3 charge eigenstates, $\Pi_{4/3}$,
$\Pi_{1/3}$ and $\Pi_{-2/3}$, with charges $4/3$, $1/3$ and $-2/3$
respectively. Since we expect them to be rather heavier than the top, if we assume all $c_{ij}$ coefficients in the couplings to have a modulus equal to 1,
their branching ratio to third generation quarks and leptons is around
94\% or greater. So they
predominantly decay as follows:
\begin{align*}
\Pi_{4/3} \rightarrow \overline{\tau}~ \overline{b}, \\
\Pi_{1/3} \rightarrow \overline{\tau}~ \overline{t} \text{ or } \Pi_{1/3} \rightarrow \overline{\nu_{\tau}}~ \overline{b}, \\
\Pi_{-2/3} \rightarrow \overline{\nu_{\tau}}~ \overline{t}.
\end{align*}

The branching ratios are quite sensitive to the $c_{ij}$ coefficients, however, so other decay modes ({\em eg.} involving second generation leptons) may be important for different values of $c_{ij}$, even if they are all still $O(1)$. The bounds and branching ratios in this section have been derived under the assumption that the modulus of all $c_{ij}$ coefficients be equal to 1, but we will comment on the impact of lifting this assumption towards the end of the section.

There will be electroweak mass splittings between the three leptoquark
states, allowing the heavier ones to decay to the lighter ones, but
these decays will be subdominant to those through the leptoquark
couplings, if the mass splittings are small.
Of the LHC leptoquark searches, dedicated searches for third
generation leptoquarks will put the strongest limits on our
leptoquarks \cite{Gripaios:2010hv}. The $\Pi_{-2/3}$ leptoquark will
decay to tops and missing energy, so stop searches, which look for the
same signature, will apply. Likewise sbottom searches will apply to $\Pi_{1/3}$.
A recent CMS search~\cite{Khachatryan:2014ura} ruled out leptoquarks
decaying wholly to $\tau$ and $b$ up to a mass of
740 GeV. This bound roughly applies to the leptoquark $\Pi_{4/3}$. This
leptoquark's branching ratio to $\tau$ and $b$ is 0.94 (over the mass
range of the search, the variation is only in higher decimal places),
so the bound on it from~\cite{Khachatryan:2014ura} is roughly 720 GeV.
Another CMS search~\cite{CMS:2014gha} puts bounds on leptoquarks decaying to either top and tau or bottom and neutrino with a combined branching ratio of 100\%. Since the $\Pi_{1/3}$ state has a combined branching ratio of 97\% to these final states, to a good approximation the results of this search should apply. This search implies a bound of 570 GeV on the mass of the $\Pi_{1/3}$, which at this mass has a branching ratio of 0.40 to top and tau.
A bound from an ATLAS stop search~\cite{Aad:2014kra} can be applied to
the remaining leptoquark state, $\Pi_{-2/3}$. In one
scenario considered in the search, the stop is presumed to decay
wholly to a top and the lightest neutralino, and a 640 GeV bound on
the mass of the stop is quoted, assuming that the neutralino is
massless. The production mechanism for the $\Pi_{-2/3}$ leptoquark is
identical to that for the stop, which is assumed in the search to be directly pair produced. Furthermore at a mass of 640 GeV, the branching ratio of $\Pi_{-2/3}$ to top and neutrino is greater than 99.5\%. We can hence take the 640 GeV bound to apply directly to the mass of the leptoquark $\Pi_{-2/3}$.
Since we are assuming small mass splittings between the charge
eigenstates in the leptoquark multiplet, a bound on the mass of one
eigenstate roughly corresponds to a bound on them all. So we can apply
the strongest of the bounds given above to the mass $M$; we therefore
conclude that $M>$ 720 GeV.

However all these bounds are found by assuming that the $O(1)$ coefficients $c_{ij}$ in the leptoquark couplings all have a modulus equal to 1. The limits can change quite a lot if this is not the case. In particular, the ratio of $\lambda_{3j}$ to	$\lambda_{2j}$ is $(c_{3j} \sqrt{m_{\tau}})/(c_{2j} \sqrt{m_{\mu}})$, ie. $\sim 4 c_{3j}/c_{2j}$. So the branching ratio to \emph{e.g.} top and muon can be larger than that to top and tau if $c_{33}/c_{23} \lesssim 0.25$. By contrast, the difference between the third and second generations of quarks is harder to overcome by changes in the $c_{ij}$ coefficients, since the hierarchy in the mixing parameters is larger. A measured bound on leptoquarks decaying to third generation quarks and second generation leptons would be a useful measurement to cover a case where $c_{33}$ were accidentally small.

\section{Conclusions}

We have argued that current anomalies in semileptonic $B$ decays seen
at LHCb are consistent with a composite Higgs model featuring an
additional, light leptoquark. This leptoquark has quantum numbers
$(\mathbf{\bar{3}},\mathbf{3},1/3)$ under the SM gauge group and couplings to the SM
fermions that are largely fixed by the partial compositeness
paradigm. We have identified a possible coset structure that contains
both the SM Higgs and the leptoquark as pseudo-Goldstone bosons of the
strong sector, which allows them to be rather lighter than the
resonances of the strong sector, and with a natural explanation of
the sizes and signs of their squared-mass parameters (bar a small,
unavoidable, residual fine-tuning in the electroweak scale).

The partial compositeness framework automatically implies lepton
non-universality in the leptoquark coupings. In this way, the
departure from unity of the value of the ratio of the branching ratio of $B \to K
\mu^+ \mu^-$ to that of $B \to K e^+ e^-$ measured at LHCb this year
can be accommodated, as can earlier anomalies in measurements of
angular observables in $B \to K^* \mu^+ \mu^-$ decays. 
The framework predicts large couplings to third-generation leptons,
hence large deviations in observables involving tau leptons or
neutrinos in the final state. These processes therefore provide a good
check for the model, but the predictions are not in conflict with
current bounds.  In our scenario, with parameters chosen to fit the
LHCb $b \to s \ell \ell$ anomalies, we predict deviations of $\sim
25\%$ from the SM value in $\mathcal{B}(B \to K \nu \overline{\nu})$,
which will be testable at the Belle II experiment. And the NA62
experiment, starting in 2015, will measure $\mathcal{B}(K^+ \to \pi^+
\nu \overline{\nu})$ to sufficient accuracy that the effects of the
model should be visible there.

We find that the model is consistent with all other flavour
constraints, however we have identified a few processes for which the
NP contributions are at or close to the current bounds. These are
$\mathcal{B}(\mu \to e \gamma)$, the mass splitting $\Delta m_{B_s}$
in $B_s$ meson mixing, and $\mathcal{B}(B^+ \to \pi^+ \mu^+
\mu^-)$. More precise measurements of these will also test the
model. It should be remembered, however, that the leptoquark
couplings, $\lambda_{ij}$, within partial compositeness are each only
predicted up to an $O(1)$ factor, $c_{ij}$. Thus predictions can only
be made to an accuracy of order one or so, and even tight constraints
could be evaded if, for a particular process, the combination of
$c_{ij}$ factors involved is accidentally small. It should be noted
that, in particular, none of the processes listed in this paragraph
get their dominant contributions from the same combination of
couplings that are involved in the LHCb $B$ decay anomalies. Thus, the
various $O(1)$ factors are not determined by fitting the anomalies. However, the fact that the framework can make predictions for a wide range of processes, due to non-zero couplings to all SM fermions, means that it is, nevertheless, falsifiable.

If the composite leptoquark is the cause of the measured discrepancies
in $B$ decays, there are three implications for the model. Firstly,
the composite sector must be strongly interacting, $g_{\rho} \sim
4\pi$. Secondly, the left handed doublet of the third quark
generation must be highly composite $\epsilon^q_3 \sim 1$. Thirdly,
the leptoquark should have a mass of around a TeV, meaning that there is
scope for its discovery at LHC13. Current LHC8 bounds, analysed in
\S~\ref{LHC}, exclude the leptoquark up to masses of
$M>720$ GeV (under an assumption on the coefficients involved in the couplings). Large third generation couplings ensure that the three
charge eigenstates of the leptoquark triplet decay mostly to third
generation quarks and leptons, so searches for third generation
leptoquarks are effective for constraining their mass.

We finally comment on the physics associated with the strong sector at the scale $m_{\rho}$. First of all let us estimate the value of that scale.
According to the discussion in \S\ref{sec:pc} we expect that $M^2 \sim \frac{\alpha_s}{4 \pi} m^2_{\rho}$.
Given that we need $M \sim 1$ TeV to explain the LHCb anomalies, we obtain $m_{\rho} \sim 10 $ TeV. With such a scale for the composite sector it has been shown \cite{KerenZur:2012fr} that the structure 
of partial compositeness is enough to suppress dangerous contributions to indirect search observables with the exception of the electron EDM and the radiative LFV decay
$\mu \to e \gamma$.\footnote{Other observables close to the current experimental sensitivity are the neutron EDM, $\epsilon_K,\epsilon' / \epsilon$ and $B \to X_s \gamma$.}
The further suppression required in these channels might be obtained by
departing from the hypothesis of lepton flavour anarchy in the
strong sector.\footnote{Various non-anarchic flavour structures have been considered in the context of partial compositeness, see for example 
\cite{Cacciapaglia:2007fw,Csaki:2008zd,Redi:2011zi,Redi:2012uj,Barbieri:2012tu,Straub:2013zca,Redi:2013pga,Matsedonskyi:2014iha}. In our framework a possible solution is to invoke an unbroken $U(1)_e$ symmetry protecting the electron family number. This will lead to vanishing leptoquark couplings with the electron field, but our phenomenological analysis is unaffected.}
Very roughly, the amount of tuning needed to accommodate the right values of the EW
scale and of the Higgs mass \cite{Matsedonskyi:2012ym,Marzocca:2012zn,Pomarol:2012qf,Redi:2012ha,Panico:2012uw} is expected to be, at best, at or below the {\em per
  cent} level, which is not much worse than the amount already required in
generic supersymmetric extensions of the SM, given current bounds.
We feel that this is a not unreasonable price to pay, given the
additional benefits of a motivated flavour paradigm and the power to
explain the LHCb anomalies.

\section{Acknowledgments}
We thank D.~Ghosh and J.~Kamenik for useful discussions. We thank R. Barbieri, D. Guadagnoli and R. Kogler for further comments. This work has
been partially supported by STFC grant ST/L000385/1 and King's
College, Cambridge.
\bibliographystyle{JHEP}
\bibliography{references}

\end{document}